\documentclass[twocolumn,multicolumn,aps,prl,amsmath,showpacs,floatfix]{revtex4}
\usepackage{graphics}%
\usepackage{graphicx}
\usepackage{dcolumn}
\usepackage{bm}
\usepackage{color}
\usepackage{latexsym}
\input{epsf}

\begin{document}

\preprint{APS/123-QED}

\title{ Composite to tilted vortex lattice transition in
Bi$_{2}$Sr$_{2}$CaCu$_{2}$O$_{8+\delta}$  in oblique fields}
\author{M. Konczykowski$^{a}$, C.J. van der Beek$^{a}$, A.E.
Koshelev$^{b}$, V. Mosser$^{c}$, M.
Dodgson$^{d}$,  and P.H. Kes$^{e}$}
\affiliation{ $^{a}$ Laboratoire des Solides Irradi\'{e}s, CNRS-UMR 7642   
\& CEA/DSM/DRECAM, Ecole Polytechnique, Palaiseau, ,France}                
\affiliation{$^{b}$ Materials Science Division, Argonne National Laboratory,
Argonne, IL 60439, U.S.A. }
\affiliation{$^{c}$ ITRON, 50 Avenue Jean Jaur\`{e}s, F-92120 
Montrouge, France}
\affiliation{$^{d}$  Department of Physics and Astronomy,   
University College, Gower Street, London WC1E 6BT, U.K.}    
\affiliation{ $^{e}$Kamerlingh Onnes Laboratory, Leiden   
University, PO Box 9506, 2300RA Leiden, the Netherlands}  

\date{\today}

\begin{abstract}

      Precision measurements of the vortex phase diagram in single
      crystals of the layered superconductor 
Bi$_{2}$Sr$_{2}$CaCu$_{2}$O$_{8+\delta}$ in oblique magnetic
      fields confirm the existence of a second phase transition, in addition to the
      usual first order vortex lattice melting line
      $H_{m}(T)$. The transition  has a strong first order 
      character, is accompanied by
      strong hysteresis, and intersects the melting line in a
      tricritical point ($H_{m}^{\perp}$, $H^{\parallel}_{cr}$).
      Its field dependence  and the changing 
      character of the melting line at the tricritical point strongly suggest
      that the ground state for magnetic fields closely aligned with
      the superconducting layers is a lattice of uniformly tilted vortex lines.

\end{abstract}

\maketitle

The first order ``vortex melting'' transition from a solid (phase-ordered)
state to a liquid
state with only short range correlations is the main feature  of the
phase diagram of vortex lines in clean, layered high-temperature
superconductors \cite{melting}.  The application of a small field component
$H^{\parallel}$, parallel to the superconducting layers, leads to a 
lattice of tilted
vortex lines that melts in a similar fashion \cite{CrossLatPRL99}. However, in the more anisotropic (layered)
compounds such as Bi$_{2}$Sr$_{2}$CaCu$_{2}$O$_{8+\delta}$, the 
depression of the perpendicular
component of the melting field $H_{m}^{\perp}$ by larger parallel
fields was interpreted as the consequence of the decomposition of the
tilted vortex lattice into a combined lattice structure of Josephson 
Vortices (JVs)
and Abrikosov-type pancake vortices (PVs) \cite{CrossLatPRL99}.
For very small field components $H^{\perp}$ perpendicular to the
layers, chain structures \cite{ChainReview} arising from the
attractive interaction of PVs with JVs were directly visualized
by Bitter decoration \cite{Bolle91,Grig95}, scanning
Hall-probe \cite{ChainReview,GrigNat01} and magneto-optical techniques
\cite{VlaskoPRB02,TokunagaPRB02}. At higher $H^{\perp} \sim H_{m}^{\perp}$,
the contribution of the JVs to the free energy of
the pancake vortex crystal results in the almost linear depression of
$H_{m}^{\perp}$ as function of the parallel field
\cite{CrossLatPRL99,OoiPRL99,KonczPhysC00,MirkovPRL01}. This
behavior in moderate $H^{\parallel}$ stops at a temperature
dependent characteristic field $H^{\parallel}_{cr}$. Even though
melting is still observed above  $H_{cr}^{\parallel}$, the variation of
$H_{m}^{\perp}$ with increasing $H^{\parallel}$ becomes much
weaker \cite{KonczPhysC00,MirkovPRL01}. Several controversial
interpretations of this changing behavior were proposed, such as 
layer decoupling
\cite{KonczPhysC00},  a commensurate transition \cite{Savel01}, and a
matching effect \cite{TokunagaPRB02A}.


In this Letter we focus on the high-temperature portion of the vortex phase
diagram in single crystalline 
Bi$_{2}$Sr$_{2}$CaCu$_{2}$O$_{8+\delta}$ in oblique fields,
which can be established precisely using the well-defined discontinuity
of the vortex density at the melting transition. 
We show that ($H_{m}^{\perp},H_{cr}^{\parallel})$
corresponds to a tricritical point in the vortex lattice phase
diagram, where the melting 
crosses a novel transition from
a composite lattice at low parallel fields, to another tilted lattice structure
at \em high \rm $H^{\parallel}$.
The experimental observation of large hysteresis suggests that this
transition is strongly first order, consistent with recent 
predictions \cite{Koshelev2006}.
  The identification of the vortex ground state at high parallel field
  as a tilted lattice structure resolves the open problem of
the apparent anisotropy factor $\gamma_{eff}$, and allows one to 
determine the enhancement
of $H_{m}^{\perp}$ by magnetic coupling. We find the temperature
dependence of $\gamma_{eff}$ to be consistent with previous 
observations \cite{SchillingPRB00,Mirkovic2002}
and in quantitative agreement with the proposed model.

\begin{figure}[t]
\begin{center}
\includegraphics[width=3.in]{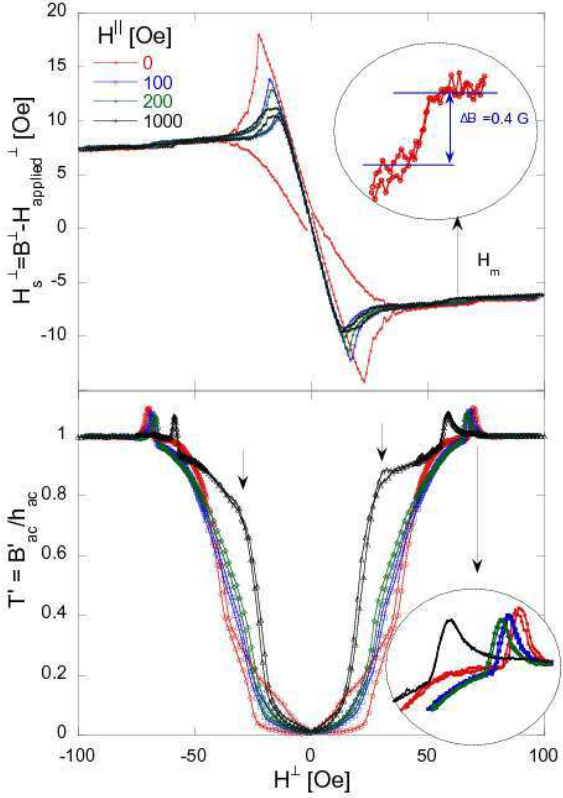}
\end{center}
\caption{(a) $dc$ local magnetization loops recorded on an as-grown
Bi$_{2}$Sr$_{2}$CaCu$_{2}$O$_{8+\protect\delta}$ single crystal ($T_{c} = 88$
K) at 75 K, as function of the magnetic field component
$H^{\perp}$ perpendicular to the superconducting layers, with the 
in-plane field $H_{\parallel}$
held constant. The inset shows a magnified view of the discontinuity at the
vortex melting transition in $H^{\parallel} = 0$. (b) the in--phase 
(screening) component of the $ac$ response
of the same crystal, recorded under the same conditions, with an $ac$ 
magnetic field of amplitude $h_{ac} = 0.8$ Oe and frequency $f = 11$ Hz applied 
along the $c$-axis. The melting transition shows up as a paramagnetic peak
(see inset), the transition from combined to tilted vortex lattice is indicated
by arrows. The $ac$ response is plotted
  as the transmittivity $T^{\prime} \equiv [B^{\prime}(f,T) - B(f,T
\ll T_{c})]/[B(f,T > T_{c}) - B(f,T \ll T_{c})]$.
}
\label{Fig:DC-AC-loops}%
\end{figure}

Experiments were performed on rectangular samples 
cut from Bi$_{2}$Sr$_{2}$CaCu$_{2}$O$_{8+\delta}$ single
crystals with different oxygen content \cite{MingLi}.
The $c$-axis component 
of the local magnetic induction $B^{\perp}({\bf r})$ was measured by micro-Hall
sensors placed on the central part of the sample. The 2D electron gas
Hall sensors were fabricated in GaAlAs heterostructures and had $8\times
8$ $\mu$m$^{2}$
active area. Results are presented in  Fig.~\ref{Fig:DC-AC-loops}(a) 
as the local magnetization $H_{s}^{\perp} \equiv
B^{\perp}-H^{\perp}$. The local $dc$ magnetization of all
crystals shows a sharp discontinuity, $\Delta B^{\perp}$, at the 
vortex melting transition, that was
tracked as function of $H^{\parallel}$ at various fixed temperatures.
The angle $\theta$ between the magnetic field and the crystalline 
$c$-axis was computer-controlled
with 0.001$^\circ$ resolution, while the field magnitude could be swept
up to 1 T using an electromagnet. Two types of magnetization loops
were measured. In the first, the magnetization is traced as function
of the $c$-axis field at constant $H^{\parallel}$; in the second,
the magnetization is measured  as function of  $H^{\parallel}$ at
constant $H^{\perp}$.

While the discontinuity in the $dc$ magnetization gives a
clear identification of the melting field,
another method \cite{MorozovPRL96}, in which the magnitude
$B(f,T)$ of the periodic part of the induction above the sample
is measured at the frequency $f$ of an
$ac$ ripple field applied perpendicularly to the sample plane, is 
more convenient and precise.
The $ac$ response is represented as the transmittivity $T^{\prime}$, 
{\em i.e.} the in-phase component
$B^{\prime}(f,T)$, normalized by the amplitude $h_{ac}$ of the $ac$ 
ripple \cite{Gilchrist93}.
The steplike feature in the $dc$ magnetization
loop at $H_{m}^{\perp}$ translates to a paramagnetic
peak in the $ac$ response, shown in Fig.~\ref{Fig:DC-AC-loops}(b) 
\cite{MorozovPRL96}.
The magnitude of this peak depends on the
ratio of $\Delta B^{\perp}$ to 
$h_{ac}$. The peak position is independent
of both the amplitude and frequency of the  $ac$ ripple. In the 
explored temperature range
(above 50 K) and at low frequency (below 27 Hz), a true paramagnetic signal
is measured. At higher frequencies or lower temperatures, flux pinning
results in the partial shielding of the $ac$ field \cite{Indenbom96}.
Nevertheless, a  peak-like feature persists at melting.

Figure~\ref{Fig:DC-AC-loops}(a) shows that at $T > 50$ K, the
application of even a small magnetic field component parallel to the
layers results in the drastic suppression of magnetic
irreversibility. 
This is expected when the geometric barrier is at the origin of flux pinning
\cite{ZeldovPRL94,Morozov97}.
%
Simultaneously, $H_{m}^{\perp}$ is depressed linearly with increasing $H^{\parallel}$. 
However, at a well-defined value $H^{\parallel}_{cr}$, the
dependence of $H_{m}^{\perp}$ on in-plane field changes
to a much slower, quadratic behavior that very well fits the anisotropic
London model, $
H_{m}(\theta)  =  H_{m0} \slash  ( \cos^{2}\theta + \sin^{2}\theta /
\gamma_{eff}^{2})^{1/2}$ \cite{BlatterPRL92}; \em i.e. \rm the perpendicular component of
the melting field
\begin{equation}
     H_{m}^{\perp}  =  \sqrt{ H_{m0}^{\perp 2} -
     \frac{H^{\parallel 2}}{\gamma_{eff}^{2}}  }
     \approx
     H_{m0}^{\perp} \left( 1 - \frac{
     H^{\parallel 2}}{ 2 \gamma_{eff}^{2} H_{m0}^{\perp 2}}   \right).
\label{Bm_theta_AL}
\end{equation}

\noindent  The characteristic field $H_{m0}$ and the
effective anisotropy parameter $\gamma_{eff}$ will be defined below.

\begin{figure}[t]
\begin{center}
\includegraphics[width=3.4in]{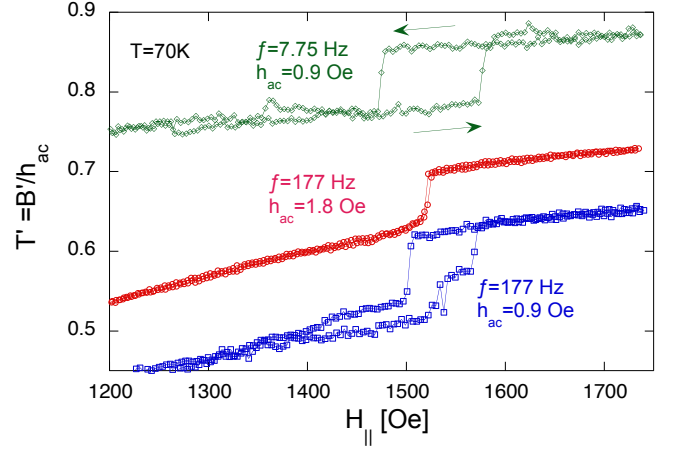}
\end{center}
\caption{Transmittivity $T^{\prime}$ of the same crystal as in
Fig.~\protect\ref{Fig:DC-AC-loops}, recorded at $T = 70$ K and constant
$H^{\perp} = 58$ Oe, as a function of $H^{\parallel}$ for
various frequencies $f$ and amplitudes $h_{ac}$ of the $ac$
ripple field. A marked hysteresis of the $ac$ screening is observed.
This hysteresis disappears when $h_{ac}$ is increased.}%
\label{Fig:Hysteresis}%
\end{figure}

In Fig.~\ref{Fig:DC-AC-loops}(b), another feature in the in-phase
component of the $ac$ response can be distinguished, at perpendicular
fields $H^{\perp}$ somewhat smaller than $H_{m}^{\perp}$.
This feature is brought out much more clearly in sweeps of the parallel
field, shown in Fig.~\ref{Fig:Hysteresis}. There is an
abrupt jump from lower to higher values of $T^{\prime}$ on
increasing $H^{\parallel}$, that only appears for parallel fields 
$H^{\parallel}
\lesssim H_{cr}^{\parallel}$. The position of the jump does not depend on $ac$
frequency. At low amplitude of the $ac$ field, a pronounced 
hysteresis of $T^{\prime}$ is
observed; this disappears if $h_{ac}$ is sufficiently increased.

The transmittivity $T^{\prime}$ is simply related to the magnitude
of the shielding current flowing in the sample in response to the
applied $ac$ magnetic field, a higher $T^{\prime}$ corresponding to
a smaller current and less screening \cite{Gilchrist93}. In the
present case, $dc$ magnetization loops point to the geometrical
barrier \cite{ZeldovPRL94} as the main source of screening. However,
increasing the $ac$ field frequency 
reduces the role of thermally activated depinning of
vortices in the crystal bulk; as a consequence, a bulk screening
current due to vortex pinning emerges \cite{Indenbom96}.  
At the frequencies of Fig.~\ref{Fig:Hysteresis}, the step in
$T^{\prime}$ is due to a discontinuous change of the magnitude of
this bulk current at the well-defined in-plane field, $H^{\parallel}
\equiv H^{\parallel}_{ct}$.  The location of $H^{\parallel}_{ct}$ does not depend on the frequency and
$h_{ac}$ which indicates \emph{a vortex phase transition in the
bulk}, from a low $H^{\parallel}$-phase with higher pinning, to a
high $H^{\parallel}$-phase with lower pinning. The hysteresis of the screening current indicates it to be first
order.

\begin{figure}[t]
\begin{center}
\includegraphics[width=3.4in]{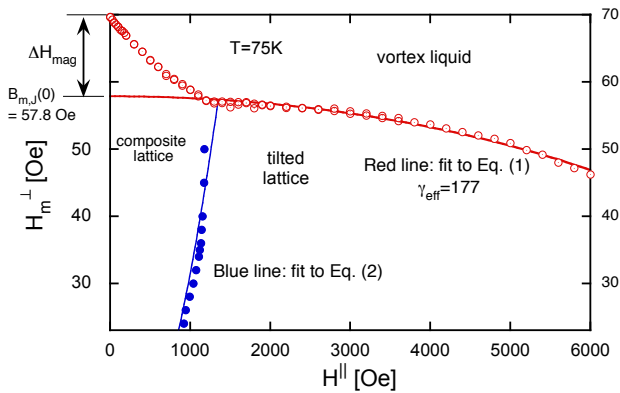}
\end{center}
\caption{Two field-component vortex-lattice phase diagram,
  with the first-order melting transition $H_{m}^{\perp}$
  ($\circ$) determined from the paramagnetic peak in $T^{\prime}$, and the
  first-order transition to the tilted PV lattice at $B_{t}$ ($\bullet$), 
  determined from the ``glitch'' in $T^{\prime}$ (Fig.~\protect\ref{Fig:DC-AC-loops} b).
  The dashed line is a fit to the composite--to--tilted lattice 
  transition, Eq.~(\protect\ref{eq:Bt}) with $C =
0.030$; the continuous line is a fit of the high-field portion of
the vortex lattice melting line to Eq.~(\protect\ref{Bm_theta_AL}).}%
\label{Fig:PhaseDiagr}%
\end{figure}

\begin{figure}[t]
\begin{center}
\includegraphics[width=3.0in]{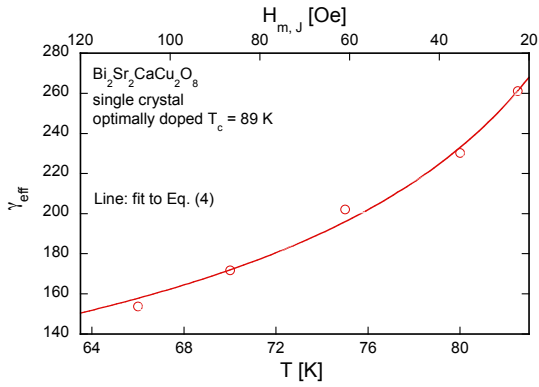}
\end{center}
\caption{Temperature dependence of the apparent anisotropy  
$\gamma_{eff}$ extracted from the fits of the melting line to
Eq.~(\protect\ref{Bm_theta_AL}).
The drawn line shows a fit to Eq.~(\protect\ref{app_anis}) with 
intrinsic $\gamma = 500$.
}%
\label{Fig:Gamma-T-HmJ}%
\end{figure}

In Fig.~\ref{Fig:PhaseDiagr} we collect, for $T = 75$ K, the
positions of the two first order transitions in a plot of $H^{\perp}$
versus $H^{\parallel}$. The usual melting field
$H_{m}^{\perp}$ of the vortices, deduced from the paramagnetic peak
in the transmittivity, shows the well-known linear decrease as function of
$H^{\parallel}$ 
\cite{CrossLatPRL99,OoiPRL99,KonczPhysC00,MirkovPRL01}, up to the
field component $H_{cr}^{\parallel}$. The field $H^{\perp}_{ct}$ of the 
first order transition revealed by the (irreversible) transmittivity rapidly 
increases with $H_{\parallel}$ and crosses the melting line
at $H_{cr}^{\parallel}$ in a tricritical point.
The same scenario is observed at all explored temperatures ($T > 50$
K),  with temperature dependent values of
$H_{cr}^{\parallel}$. The anisotropy factor $\gamma_{eff}$, extracted from
the London model fit to the high--$H^{\parallel}$ part of the melting line,
depends on temperature as well as on the oxygen content
of the Bi$_{2}$Sr$_{2}$CaCu$_{2}$O$_{8+\delta}$ crystals; it is depicted in
Fig~\ref{Fig:Gamma-T-HmJ}.
 
The low-$H^{\parallel}$ portion of the phase diagram with almost linear
$H_m^{\perp} (H^{\parallel})$ dependence has been interpreted as the
region of crossing vortex lattices of JVs and PV stacks
\cite{CrossLatPRL99}. A more accurate analysis shows that other 
``composite-lattice'' configurations compete for the ground state
in the parameter range of Bi$_{2}$Sr$_{2}$CaCu$_{2}$O$_{8+\delta}$.
These are the soliton lattice \cite{Koshelev2003}, as well as the set 
of combined lattices composed of regularly spaced rows of \em tilted \rm pancake stacks, separated by
$M$ rows of pancake stacks aligned with the $c$-axis. The latter 
type of lattice becomes favorable at smaller anisotropies and larger
$H^{\parallel}$. Moreover, if $H^{\parallel}$ is sufficiently large and the material 
anisotropy is not extremely high, a \em simple \rm tilted lattice 
($M = 0$) turns out to be the most favorable configuration. We  
interpret the experimentally observed transition as that from a composite 
to such a uniformly tilted lattice. A simple estimate for the
in-plane field at which this transition is expected, $B^{\parallel}_{ct}$, can
be obtained by comparing the ground state energies of the simplest 
($M = 1$) composite lattice and of the uniformly
tilted lattice, giving \cite{Koshelev2006}
\begin{equation}
     B^{\parallel}_{ct} \approx C \frac{\gamma}{\lambda_{ab}}
                                 \left[ B^{\perp} \Phi_{0} \slash \ln
                 \left( \frac{1.55 \sqrt{B^{\perp}\Phi_{0}}}{s 
B^{\parallel}_{ct}} \right)
                 \right]^{1/2}.
     \label{eq:Bt}
\end{equation}
Here $\Phi_{0}$ is the flux quantum, $\lambda_{ab}$ is the
$ab$--plane penetration depth, $\gamma$ is the penetration depth ratio
$\lambda_{c} / \lambda_{ab}$, and $s$ is the layer spacing.
Eq.~(\ref{eq:Bt}) gives a very good fit to the experimental transition 
line, as illustrated in
Fig.~\ref{Fig:PhaseDiagr}.

The anisotropic three-dimensional behavior (\ref{Bm_theta_AL}) of
$H_{m}^{\perp}$ for large in-plane field $B^{\parallel} > 
B_{cr}^{\parallel}$ strongly supports this interpretation.
The  $ H_{m}^{\perp}(  H^{\parallel} )$--dependence is
the direct consequence of the vanishing contribution of the magnetic
interaction between PVs to the vortex tilt stiffness in a highly
inclined tilted vortex structure. The angular dependence of the
melting field can be derived using a scaling transformation of
coordinates, $\tilde{z}=\gamma
^{2/3}z;\;\tilde{r}_{\perp}=\gamma^{-1/3}r_{\perp}$, which reduces
the larger part of the free energy to an isotropic form \cite{BlatterPRL92}.
 In scaled coordinates the magnetic
field is given by $\tilde{B} =B\gamma^{2/3}\left(
\cos^{2}\theta+\sin^{2}\theta/\gamma^{2}\right)^{1/2}$, while the
tilt angle $\tan \tilde{\theta}=\tan\theta/\gamma$. The Josephson
tilt energy of a deformed vortex line (PV stack) in scaled
coordinates,
\[
E_{J,t}=\int\frac{d\tilde{k}_{l}}{2\pi}\frac{\tilde{\varepsilon}_{1}(\tilde
{k}_{l})}{2}\tilde{k}_{l}^{2}|\delta\mathbf{\tilde{u}}(\tilde{k}_{l})|^{2},%
\]
is determined by the effective line tension
$\tilde{\varepsilon}_{1}(\tilde{k}_{l})=\tilde{\varepsilon}_{0}\ln\left(
1 \slash \tilde{r}_{\mathrm{cut}}\tilde{k}_{l} \right)$,
valid when the wave vector along the line direction, $\tilde{k}_{l}$,
is much larger than the vortex lattice zone boundary vector. Here,
$\delta\mathbf{\tilde{u}}(\tilde{k}_{l})$ is the Fourier transform of
the line deformation, and 
$\tilde{\varepsilon}_{0}=\varepsilon_{0}\gamma^{-2/3}$ with
$\varepsilon_{0}\equiv\Phi_{0}^{2}/(4\pi\lambda_{ab})^{2}$. For
near-perpendicular  fields ($\tilde{\theta} \ll 1$) the core cut-off distance
$\tilde{r}_{\mathrm{cut}}$ is determined by the
so-called thermal vortex wandering length,  $\tilde{r}_{\mathrm{cut}} \approx
\langle \mathbf{\tilde{u}}_{n,n+1}^{2} \rangle ^{1/2}
\equiv \langle ( \mathbf{\tilde{u}}_{n+1}-\mathbf{\tilde{u}}_{n} 
)^{2} \rangle ^{1/2}$,
where $\mathbf{u}_{n}$ is the position of the
PV vortex in layer $n$ \cite{Koshelev98}. 
For a {\em tilted} vortex line, $\mathbf{\tilde{u}}_{n,n+1} = 
s\tan\tilde{\theta}+\delta
\mathbf{\tilde{u}}_{n,n+1}$ consists of the average
displacement as well as random (thermal) fluctuations meaning that the core
cut-off $\tilde{r}_{\mathrm{cut}}^{2}
\approx\ s^{2}\tan^{2}\tilde{\theta}+\left\langle (\delta\mathbf{\tilde
{u}}_{n,n+1})^{2}\right\rangle  $. The melting temperature is given by
  $T_{m}=A \sqrt{
  \tilde{\varepsilon}_{1}(1/\tilde{a})\tilde{\varepsilon}_{0}} 
\tilde{a}$, with $\tilde{a}\equiv (\Phi
_{0}/\tilde{B})^{1/2}$ and $A\approx0.1$ \cite{NordborgAEKPRB99}. 
Returning to real coordinates, we obtain%
\begin{equation}
T_{m}^{2}=A^{2}(\varepsilon_{0}s)^{2}\ln\left(  \frac{C_{J}B_{sc}(\theta
)/B}{r_{0}^{2}+\tan^{2}\theta/\gamma^{2}}\right)  \frac{B_{sc}(\theta)}{B}
\label{Tm_theta}%
\end{equation}
\noindent where the numerical constant $C_{J}\approx 5$  can be 
estimated within the self-consistent harmonic
approximation, $B_{sc}(\theta)    = 
(\Phi_{0}/\gamma^{2}s^{2})/\sqrt{\cos^{2}\theta
+\gamma^{-2}\sin^{2}\theta}$, and $r_{0}^{2}   =\left\langle 
(\delta\mathbf{\tilde{u}}_{n,n+1}
)^{2}\right\rangle /(\gamma s)^{2} \approx 2A [\Phi_{0}/s^{2}
\gamma^{2}B_{m}(0) ]^{1/2}$.
Note that the angular-dependent core cutoff introduces an additional angular
dependence of melting field: $T_{m}$  no longer depends only on the 
ratio $B_{sc}(\theta)/B$.
In particular, a new angular scale appears given
by $\tan\theta=\gamma r_{0}$. In the experimental angular range 
$\tan\theta\ll\gamma,\gamma
r_{0}$, we recover Eq.~(\protect\ref{Bm_theta_AL}) with the apparent anisotropy
\begin{equation}
\gamma_{eff} \approx \gamma\left(  1+\frac{10\sqrt{B_{m}(0)\gamma^{2}s^{2}%
/\Phi_{0}}}{\ln\left[  68\sqrt{\Phi_{0}/(B_{m}(0) \gamma^{2}s^{2}%
)}\right]  }\right)  ^{-1/2} \label{app_anis}.%
\end{equation}
We note several key points. First, the
effective anisotropy $\gamma_{eff}$ is manifestly smaller than the intrinsic
$\gamma$. It increases with temperature, and is in excellent agreement
with the experimental data of Fig.\ref{Fig:Gamma-T-HmJ}, strongly
suggesting that the modified core cut--off length originating from the
tilting of the PV stacks determines the behavior of the melting line at
high parallel fields. Very similar behavior has been observed in
YBa$_{2}$Cu$_{3}$O$_{7-\delta}$ \cite{SchillingPRB00}. Next, the
prefactor $H_{m0}^{\perp} = B_{m}(0)/\mu_{0}$ in Eq.~(\ref{Bm_theta_AL}) is to be interpreted
as the hypothetical vortex melting field $H_{m,J}(\theta = 0)$ in the \em
absence \rm of the magnetic coupling between PVs. The difference $\Delta H_{mag} =
H_{m}^{\perp}(\theta = 0) - H_{m,J}(0) \approx 0.15 H_{m}^{\perp}$
between the real (experimental) melting field and this prefactor
represents the  (remarkably modest)  enhancement of the melting
field due to magnetic coupling.

Summarizing, we have established the existence of phase transition of the vortex
lattice in single crystalline Bi$_{2}$Sr$_{2}$CaCu$_{2}$O$_{8+\delta}$ in oblique fields. The
transition has a strong first order character and intersects the
usual first order vortex lattice melting line at a tricritical point
$[H_{m}^{\perp}(T),H_{ct}^{\parallel}(T)]$. For fields parallel to
the superconducting layers $H^{\parallel} < H_{cr}^{\parallel}$ the
melting line shows the signature of a composite lattice.
For $H^{\parallel}  > H_{cr}^{\parallel}$, the melting line is fully
consistent with that of a uniformly tilted lattice of PV stacks. We
thus propose that the new first order transition takes place between
the combined and the tilted vortex lattice.  For low in-plane
fields, the combined vortex lattice is stabilized by the magnetic
interaction between PV's in the same stack (vortex line). The
enhancement of the melting line in the combined lattice regime is
due to the contribution of this magnetic interaction to the vortex
line tilt stiffness.

\end{document}